\documentstyle[12pt]{article}
\begin{document}
\def\V{{\cal V}}
\def\H{{\cal H}}
\def\C{{\cal C}}
\def\K{{\cal K}}
\def\L{{\cal L}}

\pagestyle{empty}

\begin{flushright}
EFI 92-54 \\
\end{flushright}

\bigskip
\bigskip

\begin{center}

{\bf ON THE ANALYTIC STRUCTURE OF HADRONIC AMPLITUDES IN QCD}
\footnote{Work supported in part by the National Science
Foundation, grant 91-23780.}  \\

\bigskip
\bigskip

{\it Reinhard Oehme } \\

\medskip
{\it Enrico Fermi Institute and Department of Physics \\
University of Chicago, Chicago, Illinois, 60637 }

\end{center}

\bigskip

\medskip

\bigskip

\centerline{ABSTRACT}

\begin{quotation}

   Analytic properties of hadronic amplitudes are discussed
within the framework of QCD as formulated on the basis of
the BRST algebra. Local, composite fields are introduced for
hadrons. Given confinement, it is shown that hadronic amplitudes
have no thresholds or structure singularities (anomalous thresholds)
which are directly related to the underlying quark-gluon structure.
In contrast, general amplitudes of QCD must have
singularities in channels with non-zero color quantum number, which
can be related to unphysical states.

\end{quotation}

\newpage

\baselineskip 20 pt
\pagestyle{plain}
\bigskip
\smallskip

   Hadronic amplitudes, like vertex funktions and scattering
amplitudes, are physical quantities of interest in Quantum
Chromodynamics (QCD). Their analytic properties are important
from a phenomenological as well as a conceptional point of
view. Within the framework of an effective hadronic field
theory, the analytic structure of Green's functions has been explored
extensively many years ago.\footnote{We refer to [1] for references
to the older literature.} The essential input is relativistic
covariance, locality (microscopic causality) and spectral
conditions. The natural mathematical framework is the theory
of functions of several complex variables, and the main
tools for the derivation of dispersion representations
are the Edge of the Wedge Theorem and analytic completion.
The analytic structure of amplitudes, as obtained from hadronic perturbation
theory, is also used sometimes in order to find a minimal singularity
structure.

It is the purpose of this note to discuss analytic properties of
hadronic amplitudes within the framework of QCD as formulated
in covariant gauges and on the basis of the BRST algebra.
In particular, we show that there are no singularities of
hadronic amplitudes which are directly connected with the
quark-gluon structure, neither as thresholds nor as structure
singularities (anomalous thresholds). These features are a
consequence of confinement, which we understand as the exclusion
of quarks and gluons from the physical state space, so that they
cannot be produced in hadronic collisions. They form non-singlet
representations of the BRST algebra. Since unphysical
states can have components in the physical subspace, the absence of
thresholds requires a detailed proof. Furthermore, an
understanding of structure singularities in terms of poles
and ordinary thresholds is required.

We also show that, in contrast to hadronic amplitudes,
Green's functions with colored
channels must have singularities in these channels, which
generally are associated with quarks, gluons and other unphysical
excitations in the full state space with indefinite
metric.

In the following we collect some of the results which are
needed for our conclusions. These include implications of the
BRST algebra, the definition of composite hadron operators,
and the properties of structure singularities in the theory
of dispersion relations.

We consider QCD in covariant gauges defined by the gauge-fixing
term
$$
\L_{GF} = \frac{\alpha}{2}~ B\cdot B + (\partial_{\mu} B)\cdot A^{\mu},
\eqno(1)
$$
where $\alpha$ is a real parameter and $B$ the Nakanishi-Lautrup
auxiliary field. The theory is defined in a state space $\V$ of
indefinite metric, in which the constrained system is
quantized with the help of the BRST algebra. As is well known,
we can use the nilpotent, Hermitian BRST operator $Q$
in order to define a
Hilbert space which is invariant under Lorentz and equivalence
transformations in $\V$, and has vanishing ghost number [2,3]:
$$
\H = kerQ/imQ~, \\
\eqno (2)
$$
$$
kerQ = \{\Psi : ~Q\Psi = 0, ~ \Psi \in \V \}~, \\
$$
$$
imQ  = \{\Psi : ~ \Psi = Q\Phi, ~ \Phi \in \V \}~.
$$
Using $Q$, we also obtain a decomposition of the state space
$\V$ in the form
$$
\V = \V_p \oplus imQ \oplus imQ^* ~,
\eqno (3)
$$
where $\V_p$ is isomorphic to $\H$, and $kerQ = \V_p \oplus imQ$,
$kerQ^* = \V_p \oplus imQ^*$, $Q^* = \C Q\C$, with $\C$ being a
self-conjugate involution, which can be viewed as a metric matrix
in $\V$ with respect to the decomposition (3):
$$
\Psi = \left ( \begin{array} {c} \psi_1 \\ \psi_2 \\ \psi_3
\end{array}\right), ~~ \C = \left ( \begin{array}{clcr}
1 & 0 & 0  \\
0 & 0 & 1  \\
0 & 1 & 0  \\
\end{array} \right) .
\eqno(4)
$$
The indefinite inner product in $\V$ is then
$$
(\Psi ,\Phi ) = (\Psi ,\C \Phi )_{\C} =
\psi_1^* \phi_1 + \psi_2^* \phi_3 + \psi_3^* \phi_2 ~,
\eqno (5)
$$
where the subscript $\C$ denotes the ordinary inner product
with respect to the decomposition (3).

Given the completeness [4,1] of $Q$, we can assign a positive
definite metric to $\H \simeq \V_p$. All zero norm states
in $kerQ$ are contained in $imQ$, so that a state in $\H$
ca  be written symbolically as $\Psi_{\H} = \Psi_p + imQ$,
$\Psi_p \in \V_p$. In the following we will introduce
hadronic Heisenberg fields as BRST-invariant operators
in $\V$. They commute with $Q$ and leave $kerQ$ as well as
$imQ$ invariant. In our matrix notation, such operators are
of the form
$$
O = \left ( \begin{array} {clcr}
O_{11} & 0 & O_{13}   \\
O_{21} & O_{22} & O_{23}   \\
0      & 0      & O_{33}
\end{array} \right ) .
\eqno (6)
$$
We see that $O\Psi \in \H $ if $\Psi \in \H $, and with
$\Psi ,\Phi \in \H $, we have $(\Psi ,O\Phi ) = \psi_1^* O_{11}\phi_1 $,
involving only components in $\V_p$.

In weak coupling perturbation theory, the space $\H$ contains equivalence
classes of states which correspond to transverse gluons and quarks,
while all other excitations of the theory are eliminated from
$\H$ in a kinematical fashion by forming non-singlet representations
of the BRST algebra. For the full theory however, we expect that
quarks and gluons are confined, at least for a limited number of
flavors. By confinement we mean that there are hadrons in
the theory, and that in collisions of these hadrons only
hadrons can be produced. In order to realize these requirements
in the BRST formalism, we interpret confinement to mean that,
for dynamical reasons, transverse gluons and quarks are also
not elements of $\H$, which then contains only hadrons.
For less than ten flavors, one can give arguments [3,5] that transverse
gluons must be elements of non-singlet BRST representations,
and hence cannot be associated with states in $\H$. These
arguments involve renormalization group methods, and are
applicable only for zero temperature. Extensions to include
quarks are possible via the Kugo-Ojima condition, but the
necessity of this condition involves still approximations [5].
The algebraic formulation of confinement has the advantage,
that it is mathematically well defined and invariant. This is important
	for the derivation of analytic properties of hadronic
amplitudes, which relies on manifest relativistic covariance.
Other, perhaps more intuitive descriptions of confinement
may well be compatible with the general scheme we use here.

An essential element for our arguments are the spectral
conditions. Using the framework described above, we find that
only hadronic intermediate states are relevant. Typically,
we have decompositions of the type
$$
(\Psi ,XY\Phi ) = \sum_n (\Psi ,X\Psi_n )(\Psi_n ,Y\Phi ),
\eqno (7)
$$
where $X$ and $Y$ are BRST-invariant operators associated
with Heisenberg fields of hadrons, $\Psi , \Phi $ are
representatives of physical states, while $\{\Psi_n \}$ is
a complete set of states in the full state space $\V$. Now
we can show that only hadronic states contribute in the
sum (7):
$$
(\Psi, XY\Phi ) = \sum_n (\Psi, X\Psi_{\H n})(\Psi_{\H n} ,Y\Phi ) .
\eqno (8)
$$
Our assumptions imply that $\Psi ' = X\Psi \in \H $ and
$\Phi ' = Y\Phi \in \H $ if $\Psi ,\Phi \in \H $. Then we can
write $(\Psi ',\Phi ') = {\psi'_1}^* \phi'_1 = (\Psi ',P(\V_p)\Phi ')$,
with a Hermitian projection operator $P(\V_p)$, which exists
since $\V_p$ is a non-degenerate subspace of $\V$. Hence the
summation in Eq.(7) is only over the set of states $\{\Psi_{pn}\}$,
and can be rewritten in the form of Eq.(8), as may be seen using
the matrix representation [1-5]. We note that the projection operator
$P(\V_p)$ is not by itself invariant under Lorentz and equivalence
transformations in $\V$, but the way it is used here, as well as the resulting
Eq.(8), are invariant. Lorentz transformations are realized in
$\V$ by unitary mappings $U$ with $U^\dagger = \C U^* \C $. They
are BRST-invariant and of the form given in Eq.(6). Only $U_{11}$
transforms physical quantities. Transformations with $U_{11} = 1$
are equivalence transformations. They do not change physical
matrix elements. For unphysical states in $\V$, one can always
find an equivalence transformation so that a possible component
in $\V_p$ is transformed to zero [1-3].

  In order to obtain analytic properties  for hadronic
amplitudes, we need representations in terms of local hadronic
fields. The general problem of defining local Heisenberg
fields for stable, composite systems has been discussed in the
literature [6]. These fields interpolate between asymptotic fields
associated with the corresponding bound states as incoming
and outgoing free particles. To be specific, we consider
fundamental fields $\psi (x)$, and suppose that there exists
a one particle state $|k,M\rangle $ with $k^2 = M^2 $, so
that
$$
\langle 0 | B(x,\xi )|k,M\rangle \neq 0 ,
{}~~  B(x,\xi ) = T \psi (x + \xi )\overline{\psi} (x - \xi),
\eqno (9)
$$
where we ignore all inessential indices. We assume that there
exist functions $F(\xi )$ so that a local field can be defined
by the limit
$$
B_F (x) = \lim_{\xi \rightarrow 0} B(x,\xi )/ F(\xi) ,
\eqno (10)
$$
where $F(\xi )$ is as singular as the most singular matrix
element of $B(x,\xi )$, and hence as the leading term in
the corresponding operator product expansion.
The hadronic fields are not uniquely defined by Eq.(10). We
have equivalence classes which interpolate between the same
asymptotic fields and define the same S-matrix. This is an
extension of Borchers classes to state spaces with indefinite
metric.

As obtained from leading terms in the operator
product expansions, composite hadron fields should exist
in QCD. These expansions are expected to be a general property
of quantum field theories [7]. They are known to exist in lower
dimensional models. In four dimensions, the existence of local,
composite operators can be proven within the framework of
renormalizable perturbation theory. But we are interested in
non-perturbative QCD with confinement, and we must assume that
there is a class of most singular matrix elements so that
the limit in Eq.(10) defines a local field.

With composite, local fields for hadrons and the corresponding
asymptotic fields, we can obtain representations of scattering
amplitudes and vertex functions in terms of retarded and advanced
products of the Heisenberg fields by using the LSZ formalism
and weak asymptotic limits. The products may also contain
current densities as local operators constructed from
fundamental fields. The Fourier representations of hadronic
amplitudes make it possible to obtain inital analytic properties
as functions of momenta. These are consequences of the support
of the Fourier integrands, which is determined by microscopic
causality and spectral conditions. In order to derive dispersion
relations, we can then use the previously developed methods.
For certain vertex functions and forward and near-forward
scattering amplitudes, the relatively simple {\it gap method}
can be used [8]. But in general, more sophisticated mathematics
is required [9,10]. The natural mathematical framework is the theory
of functions of several complex variables. The essential tools
are the Edge of the Wedge Theorem [9] and the construction of
envelopes of holomorphy. The limitations of proofs are
generally due to the restricted ability to make use of
unitarity in order to remove unphysical anomalous thresholds [9,11].

As we have mentioned before, the spectral conditions are an
important input for the derivation of analytic properties.
Although products like $B(x,\xi )$ in Eq.(10) are not
BRST-invariant for $\xi \neq 0$, they define invariant hadron
fields in the limit $\xi \rightarrow 0$. From our previous
considerations, it then follows that intermediate state
decompositions only involve a complete set of hadron
states in $\H$. Hence, in a given channel of a hadronic
amplitude, there are no branch points or poles associated
with confined excitations, and consequently no thresholds
associated with quarks or gluons. The same is true for
{\it structure singularities} (anomalous thresholds) [9,11-13].
These are branch points, which can appear in a given channel,
but which are not directly related to the intermediate
states contributing to this channel. Rather, they are due to
poles or thresholds in crossed channels of other amplitudes, which
are related to the one under consideration via unitarity.
Structure singularities are present in the physical Riemann sheet for
loosely bound systems, but retreat into a secondary sheet
if the binding gets stronger [11]. Since unitarity connects
hadronic amplitudes only with other hadronic amplitudes,
we see that, given confinement, there are no structure
singularities associated with quarks and gluons, because
there are no corresponding ordinary poles or thresholds.

In order to illustrate the emergence of structure singularities
and their relationship to actual thresholds, we briefly
consider a vertex function for a simple model where a composite
particle of mass $M$ is interacting with observable constituents
of mass $m<M$. The structure function $F(z)$ of the vertex
satisfies a dispersion relation of the form
$$
F(z) = \frac{1}{2\pi} \int_{4m^2}^\infty ds
\frac{Im F(s+i0)}{s - z} ~,
\eqno (11)
$$
provided $M^2 < 2m^2$ , as will be seen below.
Associated with the threshold at $s = 4m^2$ is the discontinuity
$ImF(s+i0) = \rho (s+i0)G(s+i0)V(s+i0)$, $\rho (s) =
((s - 4m^2)/s)^{1/2}$ , where $V$ is the corresponding vertex for
the m-particle, and $G(s)$ is the appropriate partial wave
projection of the inelastic amplitude for $M + M \rightarrow m + m$.
Here we use the masses as symbols for the particles.
The unprojected amplitude $G(s,t)$ has t- and u-channels, which
correspond to the reaction $m + M = m + M$. Suppose the lowest
intermediate state in these channels is a single m-particle.
Then $G(s,t)$ has poles at $t = m^2$ and $u = m^2$. In the
projection $G(s)$, these poles give rise to a left hand branch cut with a
branch point at
$$
s = g(M^2) = -\frac{M^2}{m^2} (M^2 - 4m^2).
\eqno (12)
$$
{}From the discontinuity of $F(z)$ given above, it follows that
the continuation into the second Riemann sheet, associated with the
brach point at $z = 4m^2$, is given by [11]
$$
F^{II}(z) = F(z) - 2i\rho (z)\frac{G(z)V(z)}{1 + 2i\rho (z)F(z)}.
\eqno (13)
$$
We see from Eq.(13) that also $F(z)$ has a branch point at
$z = g(M^2)$, but it is in the second sheet provided
$M^2 < 2m^2$, in which case $g(M^2) < 4m^2$.
As $M^2$ increases past the point $2m^2$, one can show [11] that
the singularity at $g(M^2)$ encircles the branch point of
$F(z)$ at $z = 4m^2$, moves into the physical sheet, and then
down the real axis below $z = 4m^2$, where it may determine the
maximal range of the form factor corresponding to the M-particle
as a loosly bound system of two m-particles. For weak binding, we have
the structure singularity at $g(M = 2m - B) \approx 16mB$, as
is expected from the Fourier transform of the bound state
Schr\"{o}dinger wave function [11,12].

The example given above is for a composite system with observable
constituents, and it shows explicitly how structure singularities are
related to actual thresholds of related amplitudes.
With confined m-particles, we would neither have the
threshold at $s = 4m^2$, nor the structure singularity.
We have ignored here other singularities, like those
due to M-particles only.

In contrast to hadronic amplitudes, it is important to realize
that general Green's functions with colored channels must have
singularities in these channels, even in the presence of
confinement. For the covariant gauges discussed here, this can
be shown quite generally. As an example, let us consider the structure
function $D(k^2)$ of the gluon propagator. Using Lorentz
covariance and spectral conditions in the full state space $\V$,
we find that $D(k^2)$ is analytic in the cut $k^2$-plane.
With the help of renormalization group methods, it can then
be shown that this function must vanish for $k^2 \rightarrow
\infty$ in {\it all directions} in the complex $k^2$-plane [14].
Explicitly, we find
$$
-k^2 D(k^2,\alpha ) \simeq \frac{\alpha}{\alpha_0} + C(g,\alpha )
\left( lg\frac{k^2}
{\kappa^2} \right)^{-\gamma_{00} /\beta_0} ,
\eqno (14)
$$
where $\kappa^2 < 0$ is the normalization point, $\gamma_{00} /
\beta_0 = \frac{1}{2} (13 - \frac{2}{3} N_F )/(11 - \frac{2}{3} N_F)$,
$N_F$ = number of flavors, and $\alpha_0 = - \gamma_{00} /\gamma_{01}$.
There is no $\alpha$-dependence
in the exponent, because the gauge parameter $\alpha$
transforms in a nontrivial fashion under renormalization group
transformations, and $\alpha = 0$ is a fixed point. In the
derivation of Eq.(14) we have assumed that exact QCD
connects with the weak coupling perturbation theory asymptotically
as $g^2 \rightarrow 0$, at least as far as the leading terms
are concerned. It follows, that $D(k^2,\alpha )$ cannot be
a nontrivial entire function, at least for $0< g < g_{\infty}$,
where $g_{\infty} > 0$ is a possible first nonintegrable singularity
of $\beta^{-1} (g^2)$. There must be singularities of
$D(k^2, \alpha )$ on the positive, real $k^2$-axis,
because $C(g,\alpha )$ is not identically zero [14].
One can argue in a similar fashion for the quark propagator.

We conclude that, for QCD in covariant gauges as defined
in Eq.(1), we generally have poles and branch points in
colored channels. They can be associated with confined and
unphysical excitations. The corresponding states in $\V$ are
unobservable, because they form representations of the BRST
algebra which are not singlets, and hence are not representatives
of states in $\H$ [2,3,14].

Even though the quark-gluon structure of hadrons does not
give rise to thresholds and structure singularities of
hadronic amplitudes, we expect that it is of relevance
for discontinuities (weight functions) in dispersion
representations of these amplitudes. To the extent that
perturbative QCD is related to the weak coupling limit of the
full theory, we should see evidence of the composite structure
in regions of momentum space where the effective coupling is
small as a consequence of asymptotic freedom. In these regions,
perturbative QCD may be a reasonable tool for the approximate
computation of weight functions. The absence of quarks and
gluons from the physical space $\H$ may then be related to
the required hadronization.

The absence of structure singularities is of particular
interest for form factors of heavy quark systems, where a
constituent quark model may give a good description. The
maximal range of the Schr\"{o}dinger wave function of the
loosely bound system can be rather large, and give rise to
a large mean square radius. Nevertheless, as we have explained,
there are no structure singularities in the exact QCD
form factor, which would give branch points well below the
first hadronic threshold. But a larger mean sqare radius can
always be obtained with the help of an appropriate weight
function along the cuts starting at the hadron branch
points. Studies of this problem may be found in
recent, interesting papers by Jaffe and Mende [15],
which also contain further references.
Certainly, one would like to have a better understanding
of the weight functions within the framework of QCD.

\bigskip
\bigskip

I would like to thank Gerhard H\"{o}hler, Harry Lehmann
and Wolfhart Zimmermann for remarks or conversations
concerning the topics in this note. I am also grateful to
Paul Mende for bringing his work with R.L. Jaffe to my
attention.

\newpage
\bigskip
\bigskip

\def\pr{Phys. Rev.~}
\def\nc{Nuovo Cimento~}
\def\np{Nucl. Phys.~}
\def\ptp{Prog. Theor. Phys.~}
\def\pl{Phys. Lett.~}
\def\mpl{Mod. Phys. Lett.~}
\def\mlg{M. L. Goldberger}
\def\wz{W. Zimmermann}
\def\ro{R. Oehme}

\centerline{REFERENCES}

\begin{enumerate}

\item \ro, $\pi N$-Newsletter No.{\bf 7} (1992) 1; Fermi Inst. Rep.
EFI 92-17.

\item T. Kugo and I. Ojima, \ptp Suppl. {\bf 66} (1979) 1;\\
K. Nishijima, \np {\bf B238} (1984) 601; \\
\ro, Mod. \pl A{\bf 6} (1991) 3427.

\item \ro, \pr {\bf D42} (1990) 4209; \\
\pl {\bf B195} (1987) 60; \pl {\bf B252} (1990) 641.

\item M. Spiegelglas, \np {\bf B283} (1987) 205.

\item K. Nishijima, \ptp {\bf 75} (1986) 22; \\
K. Nishijima and Y. Okada, \ptp {\bf 72} (1984) 254.

\item \wz, \nc {\bf 10} (1958) 597; \\
K. Nishijima, \pr {\bf 111} (1958) 995.

\item K. Wilson and \wz, Comm. Math. Phys. {\bf 24} (1972) 87;
\wz, in {\it 1970 Brandeis Lectures}, eds. S. Deser, M. Grisaru and H.
Pendleton (MIT Press, Cambridge, 1971) pp.395-591; Ann. Phys. {\bf 71}
(1973) 510.

\item \ro, \nc {\bf 10} (1956) 1316; \\
\ro, in {\it Quanta}, ed. by P. Freund, C. Goebel and Y. Nambu (Univ. of
Chicago Press, Chicago, 1970) pp. 309-337.

\item H. J. Bremermann, \ro ~ and J. G. Taylor, \\
\pr {\bf 109}
(1958) 2178.

\item N. N. Bogoliubov, B. V. Medvedev and M. V. Polivanov,{\it Voprossy
Teorii
Dispersionnykh Sootnoshenii} (Fitmatgiz, Moscow, 1958); \\
N. N. Bogoliubov and D. V. Shirkov, {\it Introduction to the Theory of
Quantized Fields} (Interscience, New York, 1959);
\\ H. Lehmann, Suppl. \nc {\bf 14} (1959) 1.

\item \ro, in {\it Werner Heisenberg und die Physik unserer Zeit},\\ ed. by
Fritz Bopp (Vieweg, Braunschweig, 1961) pp. 240-259; \\
\pr {\bf 121} (1961) 1840.

\item R. Karplus, C. M. Sommerfield and F. H. Wichmann, \pr {\bf 111}
(1958) 1187; L. D. Landau, \np {\bf B13} (1959) 181;
R. E. Cutkosky, J. Math. Phys. {\bf 1} (1960) 429.

\item Y. Nambu, \nc {\bf 9} (1958) 610.

\item \ro, \pl {\bf 252} (1990) 641; ~~  \ro ~ and \wz, \pr {\bf D21}
(1980) 471, 1661.

\item R. L. Jaffe, \pl {\bf B245} (1990) 221; \\
R. L. Jaffe and Paul F. Mende,~\np {\bf B369} (1992) 189.

\end{enumerate}

\end{document}